\documentstyle[psfig,nicV]{article}
\begin{document}
%
%
\heading{%
Investigation of the $^{37}$Ar(n,p)$^{37}$Cl and $^{37}$Ar(n,$\alpha $)$%
^{34}$S reactions as a function of the neutron energy 
%
}
\par\medskip\noindent
%
\author{%
Cyriel Wagemans$^{1}$, Gert Goeminne$^{1}$, Jan Wagemans$^{2}$, Ronald Bieber$^{2}$, Marc Loiselet$^{3}$, Michel Gaelens$^{3}$, Bruno Denecke$^{2}$, Peter Geltenbort$^{4}$, Filip Kolen$^{1}$
}
\address{
Dept. of Subatomic and Radiation Physics, RUG, B-9000 Gent, Belgium
}
\address{
EC, JRC, Institute of Reference Materials and Measurements, B-2440 Geel, Belgium
}
\address{
Cyclotron Research Center, UCL, B-1348 Louvain-la-Neuve, Belgium
}
\address{
Institute Laue-Langevin, B.P.156, F-38042 Grenoble, France
}
\begin{abstract}
The energy dependent $^{37}$Ar(n,p)$^{37}$Cl and $^{37}$Ar(n,$\alpha $)$^{34}
$S reaction cross sections were determined for the first time in an
experimental effort involving three large facilities: the cyclotron of the
UCL (Louvain-la-Neuve, Belgium) where implanted $^{37}$Ar samples were
produced; the high flux reactor of the ILL (Grenoble, France) where thermal
(n,p), (n,$\alpha _0$) and (n,$\alpha _1$) cross sections of $(37\pm 4)$ b, $%
(1070\pm 80)$ b and $(290\pm 50)$ mb respectively could be determined, and
the GELINA neutron time-of-flight facility of the IRMM (Geel, Belgium) where
strong resonances were observed in the keV region.
\end{abstract}
\section{Introduction}
Neutron induced reactions on $^{37}$Ar occur in the weak component of the
s-process, where the most relevant neutron energies are situated in the keV
range. In order to calculate the Maxwellian averaged cross section in this
energy region, the reactions under investigation have to be studied with
thermal as well as with resonance neutrons. The thermal value is often used
to normalise the higher energy values and besides it is needed in the
calculation of the Maxwellian averaged cross section as a summation of the
thermal and resonance components\cite{Bao}.
\section{Sample Preparation}
Samples with a well-defined mass are of great importance to perform reliable
reaction cross section measurements, so a lot of effort was put in the
preparation and characterisation of suited samples. A detailed
description of the procedure is given in\cite{Nim}.

$^{37}$Ar atoms were produced via the $^{37}$Cl(p,n)$^{37}$Ar reaction by
bombarding a NaCl target with 30 MeV protons from a cyclotron of the UCL at
Louvain-la-Neuve (Belgium). These atoms were then ionised to the 1$^{+}$
state and implanted at 8 keV in a 20 $\mu $m thick Al-foil.
Different samples were produced, containing $10^{14}$ up to $5\times
10^{15}$ atoms. These numbers were determined at the IRMM in Geel via the
detection of the 2.6 keV KX-rays (emitted in the decay of $^{37}$Ar) with a
gas flow proportional counter.
\section{Measurements with thermal neutrons}
The experiments with thermal neutrons were performed at the end of the 87m
curved neutron guide H22 of the High Flux Reactor at the ILL in Grenoble
(France). A well-thermalised flux of about $5\times 10^8$ neutrons/cm$^2$
was available at the sample position.

\begin{figure}[t]
\centerline
{\vbox{\psfig{figure=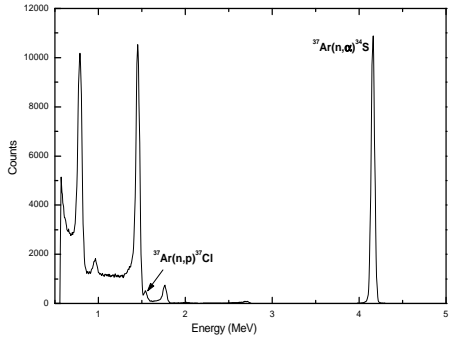,height=5cm,width=\textwidth}}}
\caption[]{\small
Measurement of the $^{37}$Ar(n,$\protect\alpha_0 $)$^{34}$%
S reaction with a sample containing $1.84\times 10^{15}$ $^{37}$%
Ar atoms.
}
\label{figure 1} 
\end{figure}

The $^{37}$Ar samples were mounted in a vacuum chamber together with suited
surface barrier detectors, placed outside the neutron beam. A typical
charged particle spectrum obtained during a 62 h neutron irradiation of one of the $%
^{37}$Ar samples is shown in figure \ref{figure 1}.
The results of these measurements are summarised in table 1.
Comparison of our results with those obtained by Ashgar et al.\cite{Ash}
shows that our values for both the (n$_{th}$,$\alpha _0$) and (n$_{th}$,p)
cross sections are about two times smaller, which indicates that the
discrepancy most likely lies in the determination of the number of $^{37}$Ar
atoms in the sample or in the neutron flux determination.

\begin{center}
\begin{tabular}{lrc}
\multicolumn{3}{l}{\textbf{Table 1.}} \\ \hline
\multicolumn{1}{c}{Reaction} & \multicolumn{1}{c}{Q-value} & Cross section
\\ 
\multicolumn{1}{c}{} & \multicolumn{1}{c}{(MeV)} & (b) \\ \hline
$^{37}$Ar(n$_{th}$,$\alpha _0$)$^{34}$S & 4.63 & $1070\pm 80$ \\ 
$^{37}$Ar(n$_{th}$,$\alpha _1$)$^{34}$S & 2.50 & $0.29\pm 0.05$ \\ 
$^{37}$Ar(n$_{th}$,$\gamma \alpha $)$^{34}$S &  & $\leq 6$ \\ 
$^{37}$Ar(n$_{th}$,p)$^{37}$Cl & 1.60 & $37\pm 4$ \\ \hline
\end{tabular}
\end{center}
\section{Measurements with resonance neutrons}
The measurements with resonance neutrons were carried out at a 9 m long
flight path of the linear accelerator GELINA of the IRMM in Geel (Belgium),
covering a neutron energy range from 10 meV up to 70 keV. The flux
determination was done via the well known $^{10}$B(n,$\alpha $)$^7$Li
reaction. An overview of the characteristics of the measurements is given in
table 2.
\begin{figure}
\centerline
{\vbox{\psfig{figure=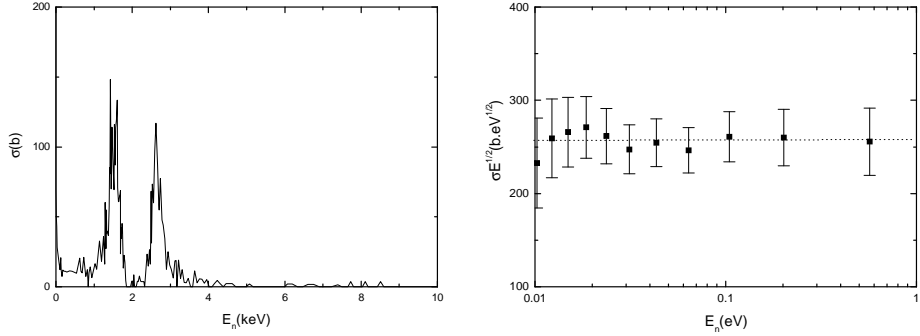,height=4.5cm,width=\textwidth}}}
\caption[]{\small
The $\sigma$ ({\it left}) and $\protect\sigma \protect\sqrt{E}$ ({\it right})
plots for the $^{37}$Ar(n,$\alpha_0$)$^{34}$S reaction
}
\label{figure 2} 
\end{figure}
In none of the three measuring cycles the (n,p), (n,$\gamma \alpha $) or (n,$%
\alpha _1$) reactions were observed, as could be expected from their small
thermal values.

\begin{center}
\begin{tabular}{p{1.2cm}p{2.2cm}p{3.3cm}p{1.75cm}p{1.4cm}}
\multicolumn{3}{l}{\textbf{Table 2.}} & \multicolumn{1}{c}{} & 
\multicolumn{1}{c}{} \\ \hline
linac frequency & detector & energy range & number of $^{37}$Ar atoms & 
irradiation time \\ \hline
100 Hz & ionisation chamber & 10 meV $\leq$ E$_n$ $\leq$ 1 keV
& $2.15\times 10^{15}$ & 150h \\ 
800 Hz & ionisation chamber & 1 eV $\leq$ E$_n$ $\leq$ 15 keV
& $1.50\times 10^{15}$ & 480h \\ 
800 Hz & surface barrier detector & 1 eV $\leq$ E$_n$ $\leq$ 
70 keV & $4.20\times 10^{15}$ & 100h \\ \hline
\end{tabular}
\end{center}

In the 100 Hz measuring campaign a 1/v shape of the $^{37}$Ar(n,$\alpha_0$)
cross section could be established (figure \ref{figure 2}).

A second measuring cycle, with the linac operating at 800 Hz, provided us
with cross section data for neutron energies up to 15 keV (figure \ref
{figure 2}). Two strong resonances were observed at 1.6 keV and at 2.5 keV,
with resonance areas of $(43\pm 9)$ b.keV and $(33\pm 7)$ b.keV.
In a third measuring cycle we used a surface barrier detector mounted in a
vacuum chamber and realised good experimental conditions up to 70 keV
neutron energy. Here, two smaller resonances were observed at 25 keV and at
40 keV. Resonance areas are in the order of 12 b.keV and 15 b.keV
respectively.
\section{Maxwellian averaged cross section}
The determination of the Maxwellian averaged cross section is based on a
formula which calculates the Maxwellian averaged cross section as a sum of
the 1/v extrapolation of the thermal value and the contributions of the
resonances\cite{Bao}: 
\begin{equation}
\left\langle \sigma \right\rangle _{kT}=\sigma _{th}\sqrt{\frac{25.3\times
10^{-6}}{kT}}+\frac 2{\sqrt{\pi }}\sum_{res}A_{res}\frac{E_{res}}{\left(
kT\right) ^2}\exp \left( -\frac{E_{res}}{kT}\right) .  \label{eq:curv}
\end{equation}
In eq. (\ref{eq:curv}) $\sigma _{th}$ is the thermal cross section value
in mb, $kT$ the stellar temperature in keV, $E_{res}$ the
resonance energy in keV and $A_{res}$ the resonance area in (mb.keV).

\begin{figure}[t]
\centerline
{\vbox{\psfig{figure=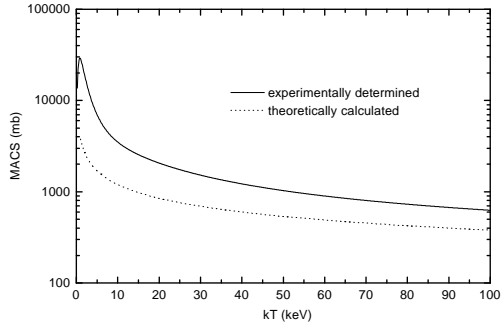,height=4.5cm,width=11cm}}}
\caption[]{\small
The Maxwellian averaged cross section for the $^{37}$Ar(n,$\alpha_0$)$^{34}$S reaction
}
\label{figure 4}
\end{figure}

Our data result in very large values for the Maxwellian averaged cross
section, e.g. 19 b at 2 keV which is 7 times larger than the theoretically
calculated one (figure \ref{figure 4}).

\section{Conclusion}
For the first time neutron induced reactions on $^{37}$Ar were performed,
covering a neutron energy range from thermal energy up to 70 keV.
Measurements with thermal neutrons were performed at the high flux reactor
of the ILL, leading to cross section values for the (n,p), (n,$\alpha _0$)
and (n,$\alpha _1$) reactions of $(37\pm 4)$ b, $(1070\pm 80)$ b and $%
(290\pm 50)$ mb respectively. For the (n,$\alpha _0$) reaction, measurements
at the neutron spectrometer GELINA of the IRMM gave evidence for a perfect 1/v
shape of the cross section in the low energy region and moreover revealed
the existence of four resonances in the region up to 70 keV. The obtained
resonance parameters combined with the thermal cross section value lead to
very large values of the Maxwellian averaged cross section.{\ }

\vfill


\begin{iapbib}{99}{
\bibitem{Ash} Ashgar M., Emsallem A., Hagberg E., Jonson B. and Tidemand-Petersson P., {\it Z. Phys} A288 (1978) 45.
\bibitem{Bao} Bao Z. and Kappeler F., {\it Atomic Data and Nuclear Data Tables} 36 (1987) 411.
\bibitem{Nim} Wagemans C., Loiselet M., Bieber R., Denecke B., Reher D. and Geltenbort P., {\it Nucl. Instr. and Meth.} A397 (1997) 22.  
}
\end{iapbib}
\end{document}